\documentclass[twocolumn]{aastex6}

\usepackage{color}

\begin{document}

\title{Does Explosive Nuclear Burning occur in Tidal Disruption Events
  of White Dwarfs by Intermediate Mass Black Holes ?}

\author{Ataru Tanikawa\altaffilmark{1,2}, Yushi
  Sato\altaffilmark{1,3}, Ken'ichi Nomoto\altaffilmark{4,5}, Keiichi
  Maeda\altaffilmark{6,4}, Naohito Nakasato\altaffilmark{7}, and Izumi
  Hachisu\altaffilmark{1}}

\altaffiltext{1}{Department of Earth Science and Astronomy, College of
  Arts and Sciences, The University of Tokyo, 3-8-1 Komaba, Meguro-ku,
  Tokyo 153-8902, Japan; tanikawa@ea.c.u-tokyo.ac.jp}
\altaffiltext{2}{RIKEN Advanced Institute for Computational Science,
  7-1-26 Minatojima-minami-machi, Chuo-ku, Kobe, Hyogo 650-0047,
  Japan}
\altaffiltext{3}{Department of Astronomy, Graduate School of Science,
  The University of Tokyo, 7-3-1 Hongo, Bunkyo-ku, Tokyo 113-0033,
  Japan}
\altaffiltext{4}{Kavli Institute for the Physics and Mathematics of
  the Universe (WPI), The University of Tokyo, 5-1-5 Kashiwanoha,
  Kashiwa, Chiba 277-8583, Japan}
\altaffiltext{5}{Hamamatsu Professor}
\altaffiltext{6}{Department of Astronomy, Kyoto University,
  Kitashirakawa-Oiwake-cho, Sakyo-ku, Kyoto 606-8502, Japan}
\altaffiltext{7}{Department of Computer Science and Engineering,
  University of Aizu, Tsuruga Ikki-machi Aizu-Wakamatsu, Fukushima
  965-8580, Japan}

\begin{abstract}

We investigate nucleosynthesis in tidal disruption events (TDEs) of
white dwarfs (WDs) by intermediate mass black holes (IMBHs). We
consider various types of WDs with different masses and compositions
by means of 3 dimensional (3D) smoothed particle hydrodynamics (SPH)
simulations. We model these WDs with different numbers of SPH
particles, $N$, from a few $10^4$ to a few $10^7$, in order to check
mass resolution convergence, where SPH simulations with $N>10^7$ (or a
space resolution of several $10^6$~cm) have unprecedentedly high
resolution in this kind of simulations. We find that nuclear reactions
become less active with increasing $N$, and that these nuclear
reactions are excited by spurious heating due to low
resolution. Moreover, we find no shock wave generation. In order to
investigate the reason for the absence of a shock wave, we
additionally perform 1 dimensional (1D) SPH and mesh-based simulations
with a space resolution ranging from $10^4$ to $10^7$~cm, using
characteristic flow structure extracted from the 3D SPH
simulations. We find shock waves in these 1D high-resolution
simulations. One of these shock waves triggers a detonation
wave. However, we have to be careful of the fact that, if the shock
wave emerged at a bit outer region, it could not trigger the
detonation wave due to low density. Note that the 1D initial
conditions lack accuracy to precisely determine where a shock wave
emerges. We need to perform 3D simulations with $\lesssim 10^6$~cm
space resolution in order to conclude that WD~TDEs become optical
transients powered by radioactive nuclei.

\end{abstract}

\keywords{hydrodynamics --- nuclear reactions, nucleosynthesis,
  abundances --- supernovae: general --- black hole physics --- white
  dwarfs}

\section{Introduction}
\label{sec:introduction}

The number of candidates for tidal disruption events (TDEs), in which
stars are tidally disrupted by a massive black hole, has been rapidly
increasing \citep[e.g.][]{2015JHEAp...7..148K}. Various stellar types
can be considered, including a main sequence star, a red giant star,
and a white dwarf (WD).  So far, several high energy transients have
been proposed to be TDEs of WDs (WD~TDEs)
\citep{2011ApJ...743..134K,2013ApJ...769...85S,2013ApJ...779...14J}.

Among various implications, finding WD~TDEs is of special importance
to address the existence of an intermediate mass black hole (IMBH) --
a black hole (BH) disrupting a WD can be an IMBH.
\footnote{WDs can also be tidally destroyed by a stellar-mass
  BH. However, this is beyond the scope of this paper. We will
  investigate it elsewhere.  } This is because a supermassive BH
(SMBH) with more than $10^5M_\odot$ just swallows a WD rather than
disrupts it \citep{1989A&A...209..103L,2004ApJ...615..855K}. WD~TDEs
can thus be an important probe to IMBHs, since a large number of
WD~TDEs may be detected with the aid of current transient surveys
(e.g., intermediate Palomar Transient Factory) and next-generation
transient surveys (e.g. the Zwicky Transient Facility and the Large
Synoptic Survey Telescope). In the future, the detections of WD~TDEs
could constrain the abundance of IMBHs in the universe, although just a
few IMBH candidates have ever been discovered by X-ray observatories,
e.g., M82~X-1 \citep{2001ApJ...547L..25M} and HLX-1
\citep{2009Natur.460...73F}.  The abundance of IMBHs will place
constraints on the formation scenarios of SMBHs
\cite[e.g.][]{1984ARA&A..22..471R}.

A WD~TDE can probably be observed as various types of transients. We
focus on the possibility of the WD~TDE observed as an optical
transient resulting from its thermonuclear explosion, although there
are many studies for possible signatures of WD~TDEs as other types of
transients
\citep{2010MNRAS.409L..25Z,2011ApJ...726...34C,2012ApJ...749..117H,2014ApJ...794....9M,2014PhRvD..90f4020C,2014ApJ...795..135E,2015ApJ...804...85S,2016ApJ...833..110I}. In
a WD~TDE, the WD is heated by compression in the direction
perpendicular to the orbital plane. Hereafter, the direction
perpendicular to the orbital plane is called the $z$-direction.  Then,
the WD could undergo explosive nuclear burning, yielding radioactive
nuclei such as $^{56}$Ni.
\citep{1989A&A...209..103L,2008CoPhC.179..184R,2009ApJ...695..404R,2012ApJ...749..117H,2015MNRAS.450.4198S}.
\cite{2008CoPhC.179..184R,2009ApJ...695..404R} studied WD~TDEs and
their nucleosynthesis signatures. Using Rosswog et al.'s data of
$0.6M_\odot$ carbon-oxygen (CO) WD, \cite{2016ApJ...819....3M}
predicted the light curve and spectrum of the WD~TDE.

\begin{figure}[ht!]
  \plotone{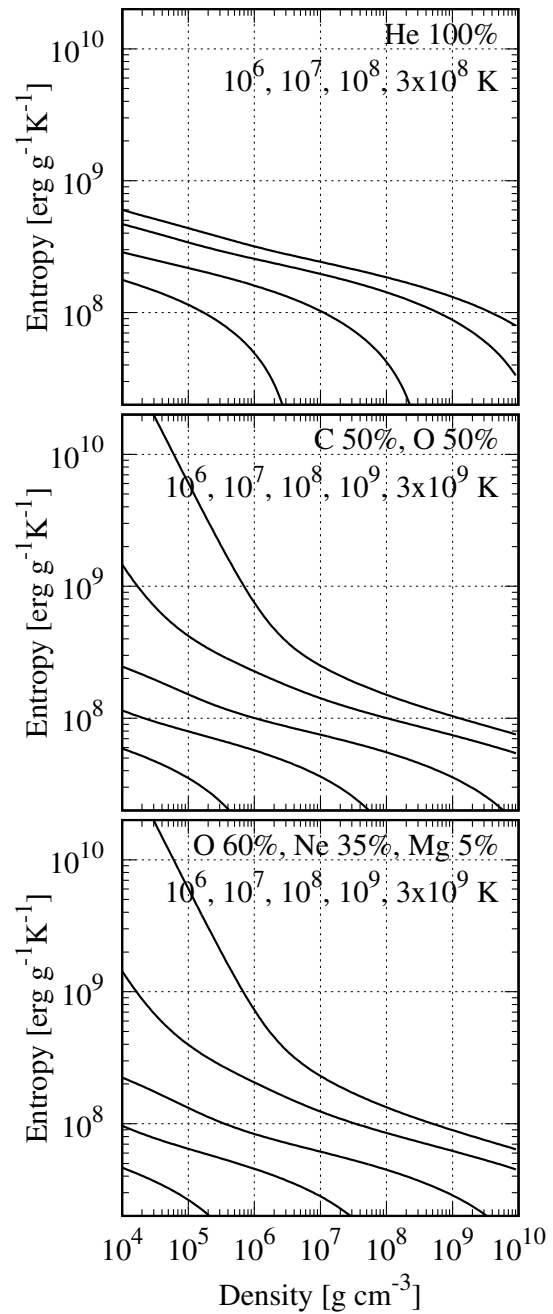}
  \caption{Density and entropy at constant temperature for materials
    consisting of He, CO, and ONeMg. The temperature shown for
    different curves is $10^6$, $10^7$, $10^8$, and $3 \times 10^8$~K
    for He, and $10^6$, $10^7$, $10^8$, $10^9$, and $3 \times 10^9$~K
    for CO and ONeMg, from bottom to top. These curves are obtained
    from the Helmholtz equation of state (EoS) described in
    section~\ref{sec:method}.
\label{fig:dens-entr}}
\end{figure}

To initiate explosive nuclear burning in a WD TDE, not only an
adiabatic compression but also a shock heating are required in order
to reach high temperature. As seen in Figure~\ref{fig:dens-entr}, a
material consisting of helium (He), CO, or oxygen-neon-magnesium
(ONeMg) needs to be adiabatically compressed by more than four orders
of magnitude to rise its temperature from $10^6$~K to $3 \times
10^8$~K (He), or $3 \times 10^9$~K (CO and ONeMg), above which the
explosive nuclear burning is triggered. However, it is difficult to
compress a WD by more than four orders of magnitude. As seen from
Figure~1 of \cite{2009ApJ...695..404R}, $\beta$, which is the ratio of
the tidal radius of the WD to a pericenter distance between the WD and
IMBH, is permitted up to about $20$. The scale height of the WD in the
$z$-direction, $z_{\textrm{min}}$, is estimated as
$z_{\textrm{min}}/R_{\textrm{wd}} \sim \beta^{-3}$ at the pericenter
\citep{1986ApJS...61..219L,2008A&A...481..259B,2013MNRAS.435.1809S},
where $R_{\textrm{wd}}$ is the original radius of the WD. Therefore,
the WD is compressed by at most a factor of $8000$. Furthermore, we
overestimate the compression of the WD in the above discussion, since
the WD is not only compressed in the $z$-direction, but also elongated
in the direction of the orbital plane.

An additional heating by the shock wave is therefore required for
initiating the nuclear reactions, however it is not clear whether
previous simulations have high resolution enough to detect such a
shock wave. \cite{2009ApJ...695..404R} have performed smoothed
particle hydrodynamics (SPH) simulations with up to $4$ million
particles. Since more than $100$ particles are aligned in the
$z$-direction, these simulations seem to resolve the shock
wave. However, the WD is elongated in the direction of the orbital
plane, and the number of particles in the $z$-direction should be
fewer than $100$, where the shock front should be perpendicular to the
$z$-direction \citep{2004ApJ...615..855K}. \cite{2012ApJ...749..117H}
have performed mesh-based simulations for a WD~TDE of $0.6M_\odot$
CO~WD encountered by a $1000M_\odot$ IMBH with $\beta=6$, and have
suggested that explosive nuclear burning is triggered at the
pericenter passage. If the original radius of the WD is $10^9$~cm, the
scale height of the WD in the $z$-direction, $z_{\textrm{min}}$ should
be $5 \times 10^6$~cm. On the other hand, the finest mesh size of
their simulation is about $10^7$~cm, larger than $z_{\textrm{min}}$.

In this paper, we perform high-resolution simulations of WD~TDEs. We
aim to investigate whether these simulations accurately follow a
thermonuclear explosion in WD~TDEs, and what is numerically required
to follow the explosion.

This paper is structured as follows. In Section~\ref{sec:method}, we
describe our simulation method. In Section~\ref{sec:result}, we
present the simulation results. In Section~\ref{sec:discussion}, we
discuss in detail the reason why the nuclear burning falsely occurs in
3D SPH simulations with low resolution. In Sections~\ref{sec:summary},
we summarize this paper.

\section{Method}
\label{sec:method}

In this section, we describe our simulation method. We overview our
SPH code in section~\ref{sec:code}. We then describe setup of 3D SPH
simulations in section~\ref{sec:3Dinit}, and of 1D simulations in
section~\ref{sec:1Dinit}.

\subsection{SPH code}
\label{sec:code}

Our SPH code solves the vanilla ice SPH equations. We adopt Wendland
$C^2$ kernel for the SPH kernel interpolation
\citep{wendland1995piecewise,2012MNRAS.425.1068D}. Our SPH code is
applicable to 3D and 1D planar geometry. The number of neighbor
particles of a given particle is about $120$ (3D) and about $5$ (1D)
unless otherwise specified. Neighbor particles are defined as
particles which are inside a sphere centered at a given particle with
its kernel-support radius, where the SPH kernel function reaches zero
at the kernel-support radius. We choose artificial viscosity proposed
by \cite{1997JCoPh.136..298M}. The artificial viscosity is dependent
on the strength of a shock wave \citep{1997JCoPh.136...41M}. The
viscosity from shear motion is suppressed by Balsara switch
\citep{1995JCoPh.121..357B}. We calculate self gravity among SPH
particles with adaptive gravitational softening
\citep{2007MNRAS.374.1347P}. We optimize the SPH and self-gravity
calculations on distributed-memory systems, using FDPS
\citep{Iwasawa:2015:FNF:2830018.2830019,2016PASJ...68...54I} and
explicit AVX instructions
\citep[e.g.][]{2012NewA...17...82T,2013NewA...19...74T}.

We use the Helmholtz equation of state (EoS) with (or without) Coulomb
corrections \citep{2000ApJS..126..501T}, which considers partially
relativistic and partially degenerate electrons including
electron-positron pairs, ions, radiation, and Coulomb interactions. We
include nuclear reactions with Aprox13 \citep{2000ApJS..129..377T}
which solves $(\alpha,p)(p,\gamma)$ and $(\gamma,p)(p,\alpha)$ links
as well as the $\alpha$-chain reaction network. The nuclear reactions
are solved implicitly if a particle has high density ($> 5 \times
10^7$~g~cm$^{-3}$) and high temperature ($> 3 \times 10^9$~K), and
otherwise solved explicitly. We can avoid overheating and overcooling
with this implicit method, even if we use a large timestep, say
$10^{-6}$~s\footnote{ \cite{2010ApJ...724..111R} said that the
  timestep should be less than $10^{-12}$~s in the explicit method, in
  order to avoid the overcooling of photo disintegration.}.  We adopt
the routines to calculate the Helmholtz EoS and Aprox13 developed by
the Center for Astrophysical Thermonuclear Flashes at the University
of Chicago.

\subsection{Setup of 3D SPH simulations}
\label{sec:3Dinit}

We follow the evolution of WD~TDEs by means of our SPH code coupled
with the nuclear reaction network. We adopt SPH modeling for the WDs,
and use a fixed potential to model the IMBH gravity.

For the IMBH potential, we adopt three kinds of models: Newtonian
potential (NP), the Paczy\'nski-Wiita (PW) potential
\citep{1980A&A....88...23P} with the modification by
\cite{2005ApJ...634.1202R}, and a generalized Newtonian potential
obtained by \cite{2013MNRAS.433.1930T}, hereafter called TR potential,
for the treatment of a BH with no spin (see also
\cite{2017arXiv170100303T} for the treatment of a spinning BH). The
IMBH is located on the coordinate origin. None of them considers the
IMBH spin.

Our initial conditions are summarized in Table~\ref{tab:initial}. We
relax the configurations of particles for these WDs in the same way as
\cite{2015ApJ...807...40T} \cite[see
  also][]{2015ApJ...807..105S,2016ApJ...821...67S}. These WDs have no
spin. We change the number of particles for the WDs, $N$, from a few
$10^4$ up to a few $10^7$. The parameter set of model CO1 is the same
as run~8 of \cite{2009ApJ...695..404R}. However, the WD radii might be
different between theirs and ours due to different ways to make
initial conditions. This might make a difference between the
pericenter distance of WDs in Rosswog's run~8 and in our model CO1,
although $\beta = 5$ in both models.

Additionally, we investigate three models. The first and second ones
are the same as models ONeMg and CO1, respectively, except without
solving a nuclear reaction network.  We call these models ``ONeMg w/o
nuc'' and ``CO1 w/o nuc''. The third one is the same as model ONeMg,
except that the number of neighbor particles is set to be proportional
to $N^{1/3}$, which is called ``ONeMg-H''. This proportionality means
that the kernel-support radii of particles are equal among different
$N$ models if the density of the particles is equal. Note the
kernel-support radii are proportional to $N^{-1/3}$ if the number of
neighbor particles is fixed.

The WD in each simulation approaches the IMBH on a parabolic orbit.
The initial separation between the WD and IMBH is set at 8 times the
tidal radius of the WD, where the tidal radii are $7.1 \times 10^9$~cm
for models CO, $1.2 \times 10^{10}$~cm for model He, and $1.5 \times
10^9$~cm for model ONeMg. Our simulations follow the evolution of
these WD~TDEs for $60$~s for models CO, $80$~s for model He, and
$10$~s for model ONeMg. At the end of the simulations, nuclear
reactions have already ceased.

\begin{deluxetable*}{llllllllllll}
\tablecaption{Summary of initial conditions. \label{tab:initial}}
\tablehead{
Model & $M_{\textrm{wd}}$ & $R_{\textrm{wd}}$ & $M_{\textrm{bh}}$ & $\beta$ &
        $R_{\textrm{p}}$  & Compositions & CC & IMBH & $N$ & Comments
}
\startdata
ONeMg & $1.2$  & $0.35$ & $100$ & $3.0$ &
        $0.52$ & $^{16}$O $60$\% $^{20}$Ne $35$\% $^{24}$Mg $5$\%  &
        w/  & PW & up to $13$M \\
CO1   & $0.6$  & $0.75$ & $500$ & $5.0$ &
        $1.4$  & $^{12}$C $50$\% $^{16}$O $50$\% &
        w/o & NP  & up to $6.3$M & Rosswog's run~8 \\
CO2   & $0.6$  & $0.75$ & $500$ & $5.0$ &
        $1.4$  & $^{12}$C $50$\% $^{16}$O $50$\% &
        w/  & TR & up to $6.3$M \\
CO3   & $0.6$  & $0.75$ & $500$ & $5.0$ &
        $1.4$  & $^{12}$C $50$\% $^{16}$O $50$\% &
        w/  & PW & up to $25$M \\
He    & $0.3$  & $1.00$ & $500$ & $5.0$ &
        $2.4$  & $^{4}$He $100$\% &
        w/  & PW & up to $25$M \\ \enddata
\tablecomments{$M_{\textrm{wd}}$ and $M_{\textrm{bh}}$ are,
  respectively, the masses of a WD and IMBH in units of
  $M_\odot$. $R_{\textrm{wd}}$ and $R_{\textrm{p}}$ are, respectively,
  the WD radius and pericenter distance in units of $10^9$~cm. $\beta$
  is the ratio of the WD tidal radius to the pericenter distance. CC
  is an abbreviation for Coulomb correction. IMBH indicates the
  gravitational potential of the IMBH, where NP, PW, and TR are
  abbreviations for Newtonian, the Paczy\'nski-Wiita, and the
  Tejeda-Rosswog potentials, respectively.}
\end{deluxetable*}

\subsection{Setup of 1D simulations}
\label{sec:1Dinit}

We construct 1D initial conditions, by extracting 1D flow structure in
the $z$-direction from the results of model ONeMg w/o nuc in the 3D
SPH simulations. Details are described in section~\ref{sec:1D}. Using
these initial conditions, we perform 1D SPH simulations, and FLASH
simulations.

Since the 1D SPH numerical methods are the same as the 3D SPH
simulations except for the dimension, we overview the FLASH
simulations here. The FLASH code \citep{2000ApJS..131..273F} is an
Eulerian code. We adopt uniform mesh, although the FLASH code supports
adaptive mesh refinement. We choose the piecewise parabolic method
\citep{1984JCoPh..54..174C} for the gas hydrodynamic solver. We use
the Helmholtz EoS and nuclear reactions with Aprox13, which are also
used for our SPH simulations. The timestep is chosen to be the minimum
value of the hydrodynamics timestep and nuclear reaction timestep. The
hydrodynamics timestep is 10~\% of the Courant-Friedrichs-Lewy number,
and the nuclear reaction timestep is 1~\% of the ratio of the specific
internal energy to the specific nuclear energy-generation rate.

We describe the 1D initial conditions in section~\ref{sec:1D}. Note
that we make the 1D initial conditions, based on the results of the 3D
SPH simulations (which is described in section~\ref{sec:3D}).

\section{Results}
\label{sec:result}

In section~\ref{sec:3D}, we present the results of our 3D SPH
simulations. The results show that the amount of the materials
experiencing explosive nuclear burning is decreased with increasing
$N$, and that the explosive nuclear burning found in these simulations
is a numerical artifact due to low resolution, not physical
effects. Furthermore, we do not find shock waves in our 3D SPH
simulations. The absence of the shock waves may be also due to low
mass resolution, even if $N>10^7$. In order to fix this problem, we
perform 1D simulations with high resolution, by extracting data from
the 3D SPH simulations. We show the results in section~\ref{sec:1D}.

\subsection{3D simulations}
\label{sec:3D}

\begin{figure}[ht!]
  \plotone{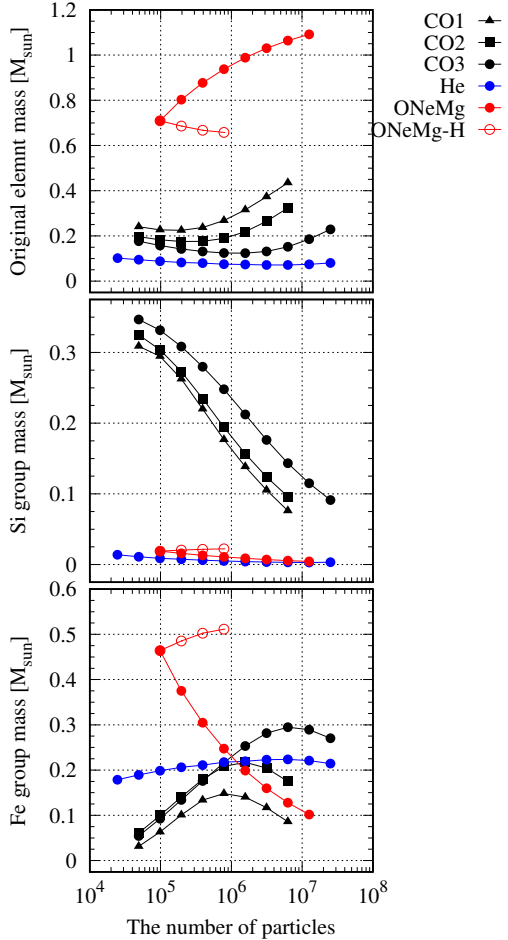}
  \caption{Masses of the original elements, Si group ($^{28}$Si,
    $^{32}$S, $^{36}$Ar, $^{40}$Ca, and $^{44}$Ti) and Fe group
    ($^{48}$Cr, $^{52}$Fe, and $^{56}$Ni) as a function of $N$ at the
    time just after nuclear reactions have ceased. Mass accreted by an
    IMBH is negligible. \label{fig:n-element}}
\end{figure}

In Figure~\ref{fig:n-element}, we summarize the masses of the
original, Si group, and Fe group elements at the time just after
nuclear reactions have ceased. Mass accreted by an IMBH at this time
is negligible. The difference in the nuclear burning products does not
come from mass accretion, but comes from the nuclear reactions.

In model ONeMg, the amount of original (unburned) elements increases,
and the amounts of Si and Fe group elements decrease, with increasing
$N$. In models CO, the amount of Si group elements decreases
monotonically with increasing $N$. Although the amount of the original
elements decreases first, it finally increases from some $N$. The
dependence of Fe group on $N$ is opposite to that of the original
elements. In model He, the dependence of each element on $N$ is
similar to those of models CO, although the dependence is
smaller. Overall, the amount of unburned materials increases at large
$N$ (say $N>10^6$) regardless of the WD masses and compositions.

\begin{figure*}[ht!]
  \plotone{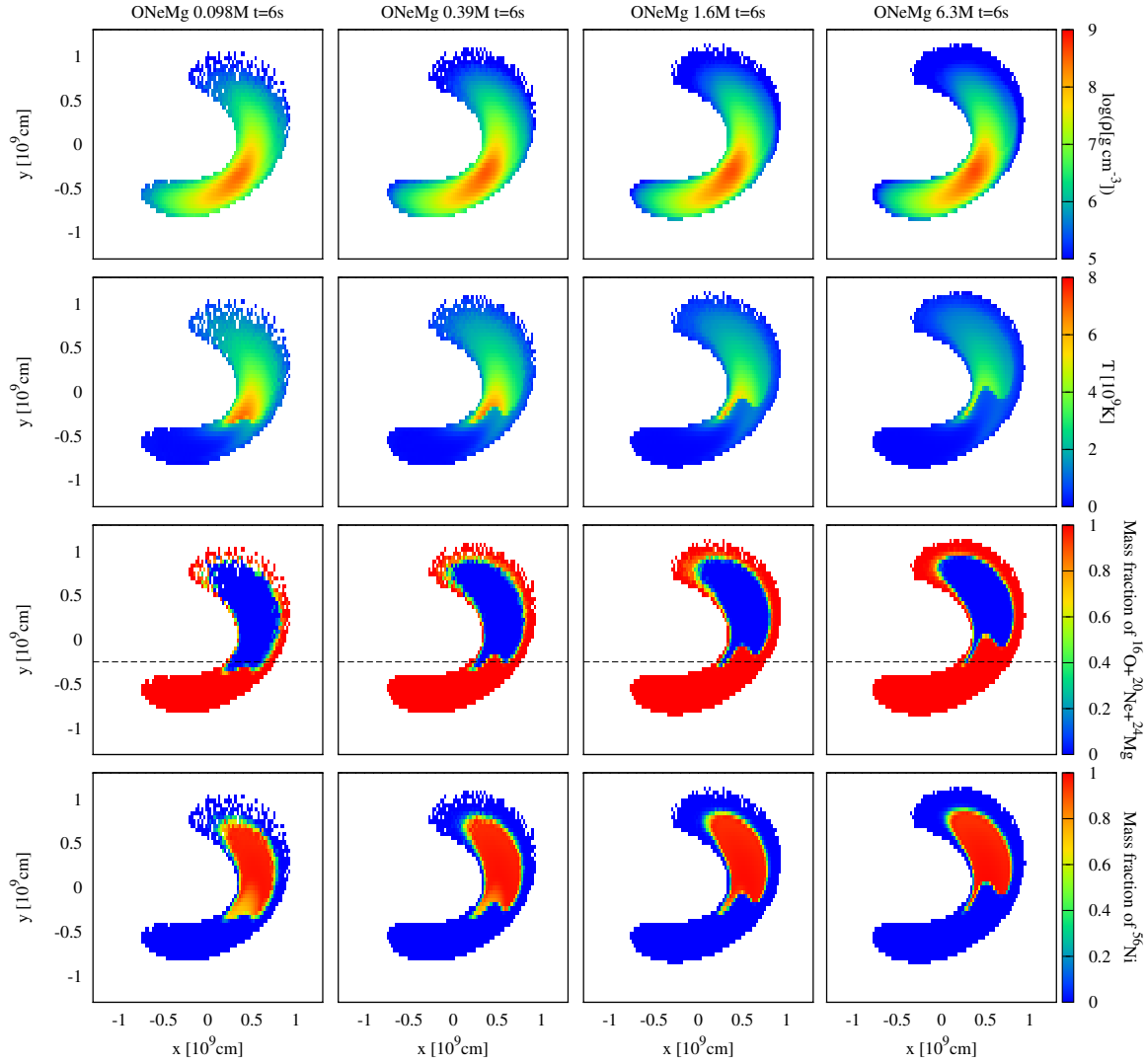}
  \caption{Density, temperature, and mass fractions of original
    elements and $^{56}$Ni on the orbital plane in model ONeMg at
    $t=6$~s, when nuclear reactions are going on. From left to right,
    $N$ is $0.098$M, $0.39$M, $1.6$M, and $6.3$M. The dashed lines
    indicate cross sections drawn in
    Figure~\ref{fig:xzplane_onemg}. \label{fig:xyplane_onemg}}
\end{figure*}

Since the nucleosynthesis of model ONeMg indicates the strongest
dependence on $N$, we show thermodynamical quantities of model ONeMg
in Figure~\ref{fig:xyplane_onemg}. At the time, nuclear reactions are
still in progress. The density of particles is not sensitive to $N$
(see the top panels). On the other hand, their temperature strongly
depends on $N$ (see the second top panels). High-temperature (say $> 2
\times 10^9$~K) region becomes smaller as $N$ increases. As the
high-temperature region becomes smaller, the amount of the unburned
materials increases (see the second bottom panels), and the amount of
$^{56}$Ni decreases (see the bottom panels), since the materials
consisting of ONeMg experience explosive nuclear burning at a
temperature above $2.5 \times 10^9$~K.

\begin{figure*}[ht!]
  \plotone{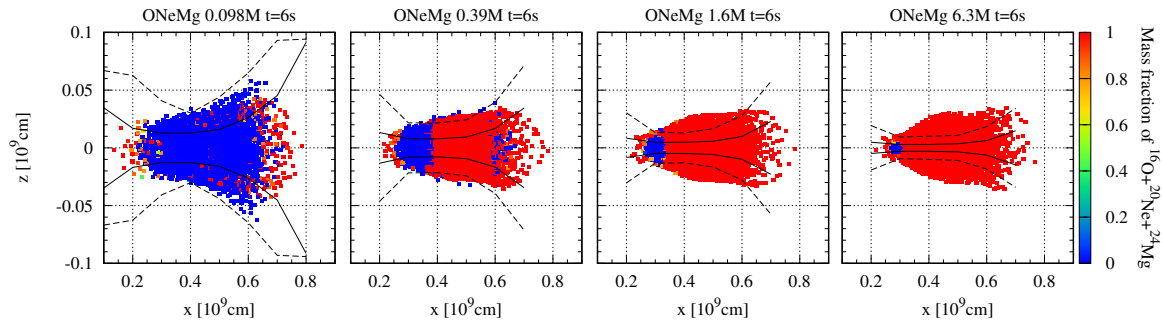}
  \caption{Mass fractions of original elements on a plane
    perpendicular to the orbital plane in model ONeMg at $t=6$~s. The
    plane is indicated by the dashed lines in
    Figure~\ref{fig:xyplane_onemg}. The width of two black solid
    (dashed) curves in the $z$-direction is equal to twice the minimum
    (maximum) value of the kernel-support radii among the particles at
    the $x$-coordinate. From left to right, $N$ is $0.098$M, $0.39$M,
    $1.6$M, and $6.3$M. \label{fig:xzplane_onemg}}
\end{figure*}

Figure~\ref{fig:xzplane_onemg} shows the structure of model ONeMg on a
plane perpendicular to the orbital plane. Although most of the
materials are burned out for $N=0.098$M, the kernel-support radius is
comparable to the scale height of the structure in the
$z$-direction. For $N=6.3$M, only a small portion of the WD
experiences nuclear reactions. However, the scale height of the small
portion is comparable to the kernel-support radius. A large portion of
the WD, where the scale height is much larger than the kernel-support
radius, is not burned.

Overall, the kernel-support radius becomes sufficiently smaller than
the scale height as $N$ increases. Instead, nuclear reactions become
inactive. The nuclear reactions occur only when the SPH simulations do
not resolve the scale height of the structure in the
$z$-direction. Therefore, we conclude that nuclear reactions
artificially occur due to numerical effects, not due to physical
effects, even if $N$ is a few $10^7$.

\begin{figure*}[ht!]
  \plotone{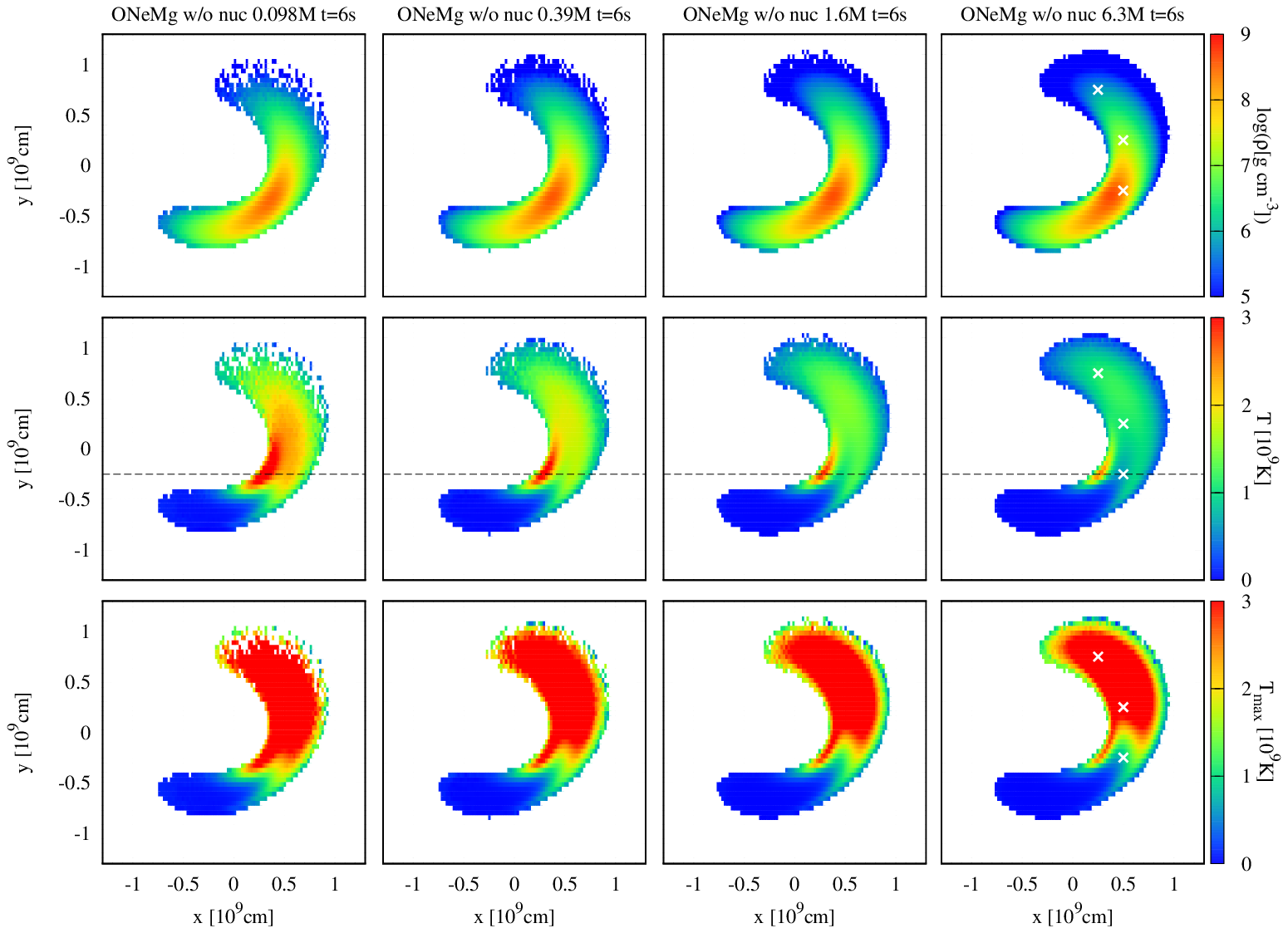}
  \caption{Density and temperature at $t=6$~s, and the maximum
    temperature each particle experiences from $t=0$~s to $t=6$~s on
    the orbital plane in model ONeMg w/o nuc. From left to right, $N$
    is $0.098$M, $0.39$M, $1.6$M, and $6.3$M. The dashed lines
    indicate cross sections drawn in
    Figure~\ref{fig:xzplane_onemg.wonuc}. The white cross points
    indicate the locations of particles for 1D SPH simulations in
    section~\ref{sec:1D}. \label{fig:xyplane_onemg.wonuc}}
\end{figure*}

We investigate model ONeMg w/o nuc in order to confirm that nuclear
reactions falsely occur, not due to any errors in our nuclear reaction
network, but due to the resolution effect of our SPH
simulations. Figure~\ref{fig:xyplane_onemg.wonuc} shows
thermodynamical quantities of model ONeMg w/o nuc on the orbital
plane. The density distributions are similar to those in model
ONeMg. The temperature distributions are different from those in model
ONeMg. Overall, the temperature in this model is lower than in model
ONeMg, since the nuclear reaction network is turned off.

The distribution of the maximum temperature each particle experiences
from $t=0$~s to $t=6$~s is a good indicator for where explosive
nuclear burning would occur from $t=0$~s to $t=6$~s if the nuclear
reaction network was turned on. Materials which reach a temperature of
$2.5 \times 10^9$~K experience explosive nuclear burning if their
density is more than several $10^5$~g~cm$^{-3}$. Therefore, explosive
nuclear burning would occur in red regions. These red regions are
almost coincident with the regions where original components are
burned out and a large amount of $^{56}$Ni is produced in model ONeMg
(see Figure~\ref{fig:xyplane_onemg}). Moreover, these red regions
shrink as $N$ increases, which is consistent with the results in model
ONeMg.

\begin{figure*}[ht!]
  \plotone{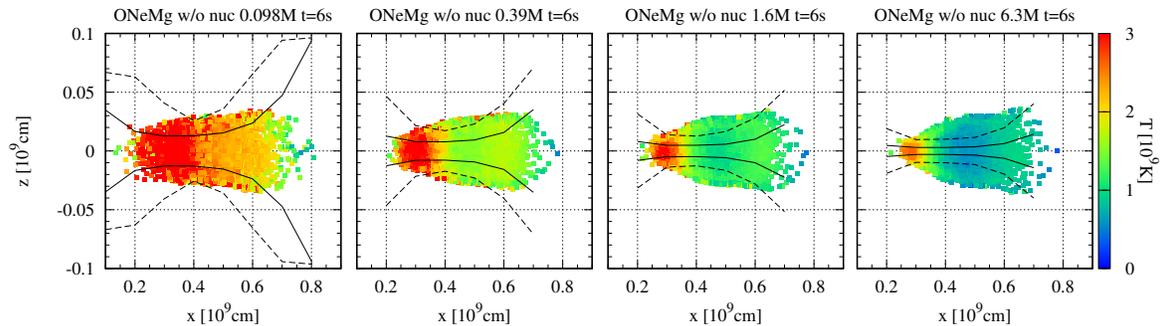}
  \caption{Temperature on a plane perpendicular to the orbital plane
    in model ONeMg w/o nuc at $t=6$~s. The plane is indicated by the
    dashed lines in Figure~\ref{fig:xyplane_onemg.wonuc}. The width of
    two black solid (dashed) curves in the $z$-direction is equal to
    twice the minimum (maximum) value of the kernel-support radii
    among the particles at the $x$-coordinate. From left to right, $N$
    is $0.098$M, $0.39$M, $1.6$M, and
    $6.3$M. \label{fig:xzplane_onemg.wonuc}}
\end{figure*}

Figure~\ref{fig:xzplane_onemg.wonuc} shows the temperature
distributions on cross sections in model ONeMg w/o nuc. The locations
of the cross sections are the same as those in model ONeMg. In
$N=0.098$M, temperature is more than $2.5 \times 10^9$~K almost in the
region. Therefore, explosive nuclear burning would occur. The
kernel-support radii are comparable to the scale heights over all the
range of $x$-coordinate. As $N$ increases, the kernel-support radii
becomes smaller, and the temperature becomes lower. In $N=6.3$M, the
kernel-support radii at $x=0.5 \times 10^9$~cm are smaller than the
scale height by a factor of a few, and the temperature is lower than
$10^9$~K. Explosive nuclear burning would not occur there. At
$x=0.3\times 10^9$~cm, the kernel-support radii are comparable to the
scale height, and the temperature is more than $2.5 \times
10^9$~K. Explosive nuclear burning would occur.

These results are consistent with those in model ONeMg. Therefore,
explosive nuclear burning in model ONeMg occurs when SPH simulations
fail to resolve the scale heights of WD~TDEs. In other words, nuclear
reactions seen in the low-resolution runs are a numerical artifact due
to spurious heating of SPH simulation.

We comment the reasons why a leading part of the WD remains high
temperature with increasing $N$, and why a trailing part of the WD
gets lower temperature with increasing $N$ (see the bottom panels of
Figure~\ref{fig:xyplane_onemg.wonuc}). At the pericenter passage, the
leading part is closer to an IMBH than the trailing part. The leading
part has smaller scale height than the trailing part. Our SPH
simulations almost resolve the scale height of the trailing part
unless $N<10^5$ (see Figure~\ref{fig:xzplane_onemg.wonuc}). However,
our SPH simulations fail to resolve the scale height of the leading
part, even if $N>10^7$. Consequently, the leading part gets high
temperature, even if $N>10^7$.

There is another evidence that the nuclear reactions occur only when
the kernel-support radii of particles are comparable to the scale
height of the structure. In model ONeMg-H, the kernel-support radii do
not become smaller by design even if $N$
increases. Figure~\ref{fig:n-element} shows that the nucleosynthesis
of model ONeMg-H is much less sensitive to $N$ than that of model
ONeMg.

Finally, we search for shock waves in model ONeMg w/o nuc, focusing on
regions where the SPH kernel is smaller than the scale height. However
we find no shock wave. If the shock waves were generated, temperature
would increase discontinuously in the $z$-direction. However, we do
not find such a temperature structure, for example in the right panel
of Figure~\ref{fig:xzplane_onemg.wonuc}.

\subsection{1D simulations}
\label{sec:1D}

We do not find shock waves in model ONeMg w/o nuc, despite the fact
that SPH kernel-support radii are smaller than the scale height (see
the right panel of Figure~\ref{fig:xzplane_onemg.wonuc}). We suspect
that the absence of the shock waves is due to the low resolution of
our simulations, even with $N > 10^7$. In order to clarify the issue,
we perform 1D planar simulations with high resolution, by extracting
local flow structures from model ONeMg w/o nuc with $N=6.3$M. For the
1D planar simulations, we use our 1D SPH code and FLASH code.

We make initial conditions for the 1D simulations as follows. We
re-perform 3D SPH simulations in order to record physical quantities
of particles at every timestep. The recorded particles are located at
regions pointed by white crosses in the rightmost panel of
Figure~\ref{fig:xyplane_onemg.wonuc}. From bottom to top, particles
get higher temperature. So, we call these regions ``low-T region'',
``middle-T region'', and ``high-T region'' from bottom to top.

From the recorded data in each region, we extract physical quantities
at the moment when $z$-direction relative velocities between the
outermost particle and a particle on the orbital plane reaches a
peak. Based on distribution of density and velocity of the extracted
data along the $z$-direction, we map particles for 1D simulations, and
assign physical quantities for 1D mesh. We fix the temperature to be
$10^7$~K. For the FLASH simulations, we put atmosphere with a constant
density of $1$~g~cm$^{-3}$ if there is no WD material. The total mass
of atmosphere is negligible, ($\lesssim 0.0001$~\% of the total mass
of WD materials).

We describe the recipe of the 1D SPH and FLASH simulations in the
following. We consider only hydrodynamics for both 1D SPH and FLASH
simulations. We do not solve nuclear reactions unless otherwise
specified. Furthermore, neither do we consider self gravity among
particles, nor the IMBH gravity. The number of particles we use is
$N=10$, $10^3$, and $10^4$ for the 1D SPH simulations. The number of
grids we use is $N_{\textrm{g}}=40$, $1600$ and $3200$. Note that the
space resolution in the largest $N$ and $N_{\textrm{g}}$ is $\sim
10^4$~cm and $\sim 10^5$~cm for the 1D SPH and FLASH simulations,
respectively. In some cases of $N_{\textrm{g}}=3200$, we perform
additional simulations, solving hydrodynamics coupled with the nuclear
reaction network.

In the cases of the low-T and high-T regions, we perform additional 1D
SPH simulations with $N=10^4$ and FLASH simulations with
$N_{\textrm{g}}=3200$. In these simulations, we change the initial
$z$-direction velocity by increasing the $z$-direction velocity for
the low-T region, and by decreasing the $z$-direction velocity for the
high-T region, so that density at $z=0$ in 1D simulations is equal to
that in 3D simulations when materials are most compressed. The reason
for this additional setups is the following. We underestimate and
overestimate the compression of materials in the low-T and high-T
regions, respectively, without the correction of the $z$-direction
velocity (hereafter, $v_{\textrm{z}}$ correction). These underestimate
and overestimate come from the facts that we do not consider an IMBH
and self gravity, and that materials are not elongated in the
direction of the orbital plane in these 1D simulations, respectively.

\begin{figure*}[ht!]
  \plotone{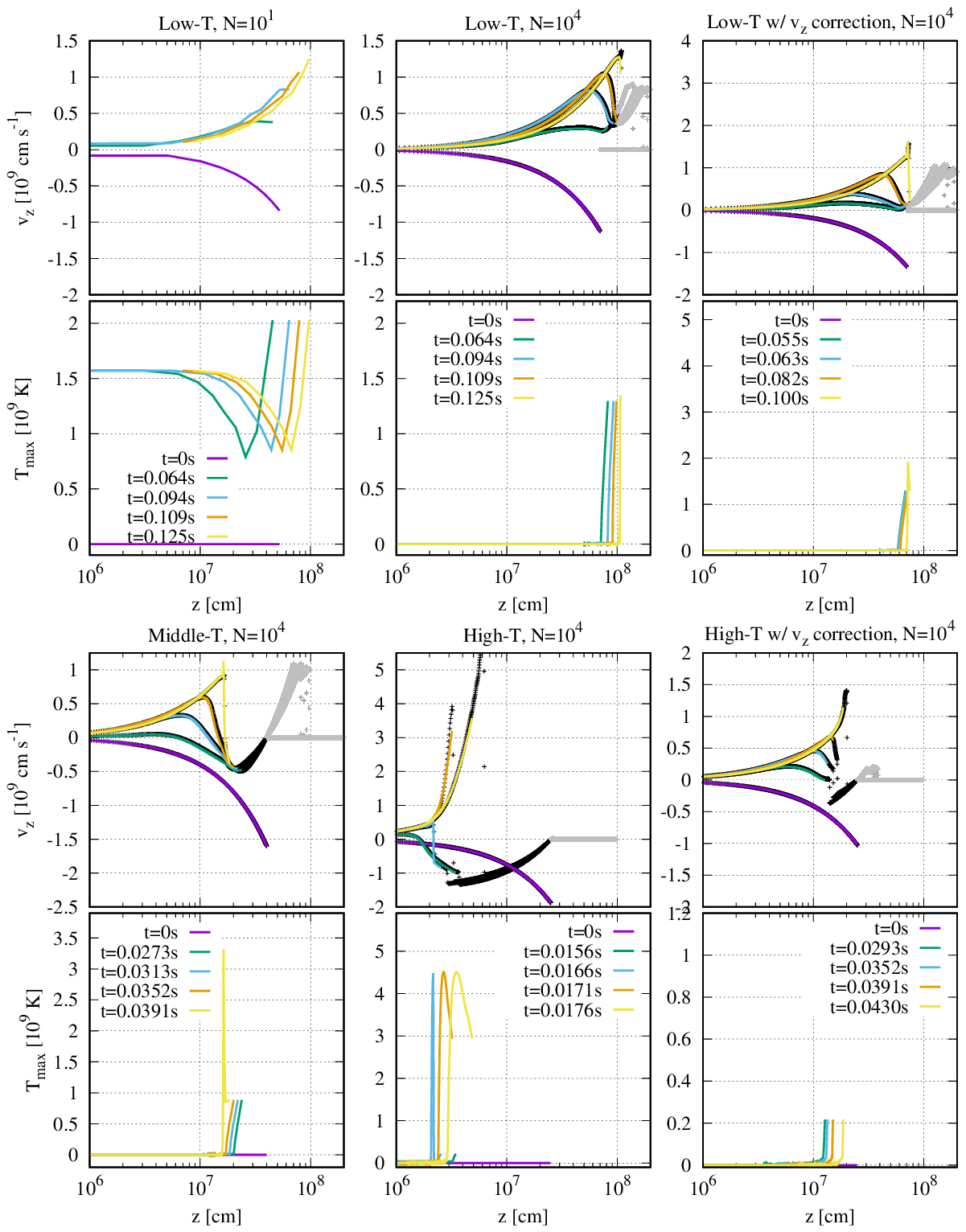}
  \caption{Time evolution of $z$-direction velocity and the maximum
    temperature each particle experiences from the initial time to the
    indicated time. Note that the initial time (i.e. $t=0$~s) is set
    to be the time when these 1D simulations are started.  The initial
    conditions and the number of particles are indicated at the top of
    the panels. Black and gray plus signs in the panels of the
    $z$-direction velocity indicate WD materials and atmosphere,
    respectively, in FLASH simulations with
    $N_{\textrm{g}}=3200$. \label{fig:sim1d}}
\end{figure*}

\begin{figure}[ht!]
  \plotone{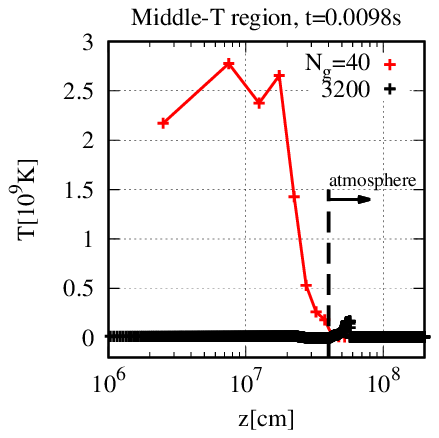}
  \caption{Temperature profile at $t=0.0098$~s in FLASH simulations
    with $N_{\textrm{g}}=40$ and $3200$ in the case of the middle-T
    region. Note that there is atmosphere at $z > 4 \times
    10^7$~cm. \label{fig:flash}}
\end{figure}

Figure~\ref{fig:sim1d} shows the results of the 1D simulations. We set
the initial time (i.e. $t=0$~s) to be the time when these 1D
simulations are started. For the low-T region with $N=10$, the
temperature on the orbital plane becomes high due to the spurious
heating. Actually, the spurious heating is seen in the results of
low-resolution simulations, i.e. the 1D SPH simulations with $N=10$
and the FLASH simulations with $N_{\textrm{g}}=40$ (see
Figure~\ref{fig:flash}), regardless of the regions. This is consistent
with our 3D SPH simulations in section~\ref{sec:3D}.

As seen in Figure~\ref{fig:sim1d}, the results of the 1D SPH
simulations are in good agreement with those of FLASH simulations, if
the simulations have high resolution, i.e. the 1D SPH simulations with
$N=10^4$ and the FLASH simulations with $N_{\textrm{g}}=3200$. For
this comparison, atmosphere should be ignored. Furthermore, we
investigate the convergence of these results against $N$ and
$N_{\textrm{g}}$. Comparing the results of $N=10^3$ and $10^4$ in the
1D SPH simulations, and those of $N_{\textrm{g}}=1600$ and $3200$ in
the FLASH simulations, we find these results are quite
similar. Therefore, these results are converged against $N$ and
$N_{\textrm{g}}$.

\begin{figure}[ht!]
  \plotone{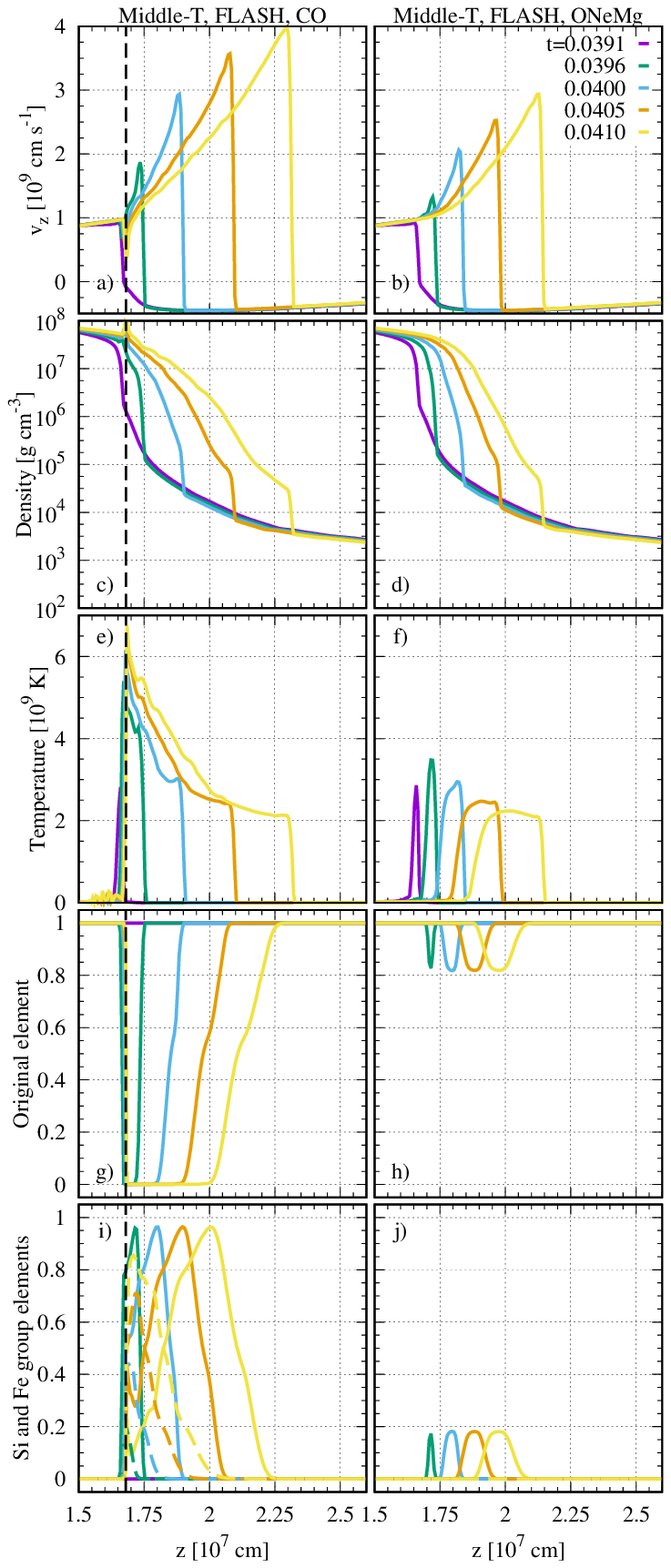}
  \caption{Time evolution of $z$-direction velocity, density,
    temperature, and mass fractions of nuclear elements in the
    middle-T region, when nuclear reactions are considered. The
    initial compositions are CO (left) and ONeMg (right). The number
    of meshes is $N_{\textrm{g}}=3200$. The time of the same colored
    curve in each panel is the same as in panel (b). In panels (i) and
    (j), solid and dashed colored curves indicate mass fractions of Si
    and Fe group elements, respectively. Note that the mass fractions
    of Fe group elements are always zero in panel (j). The black
    dashed lines in the left panels show the location of a detonation
    wave. \label{fig:det_extT}}
\end{figure}

As seen in the high-resolution results of Figure~\ref{fig:sim1d}, in
all the regions, shock waves are observed. In order to investigate
whether these shock waves trigger detonation waves, we perform FLASH
simulations coupled with the nuclear reaction network for the middle-T
region. We consider two cases where materials consist of CO and ONeMg.

Fluid motion for the CO and ONeMg cases is the same as fluid motion
for the case without considering nuclear reactions until a shock wave
emerges at $t \sim 0.0391$~s. The middle-T panels of
Figure~\ref{fig:sim1d} show the fluid motion in the case without
considering nuclear reactions. At $t=0$~s, all the materials
shrink. At $t \sim 0.0273$~s, materials at $z \sim 0$~cm bounce
back. At $t \sim 0.0391$~s, a shock wave is generated at $z \sim 1.7
\times 10^7$~cm. At this time, materials at $z \lesssim 1.7 \times
10^7$~cm are expanding, and the rest are shrinking. The fraction of
the expanding materials is $99.96$~\%. We emphasize that the shrinking
materials (i.e. $0.04$~\% of materials) are not atmosphere.

\begin{figure}[ht!]
  \plotone{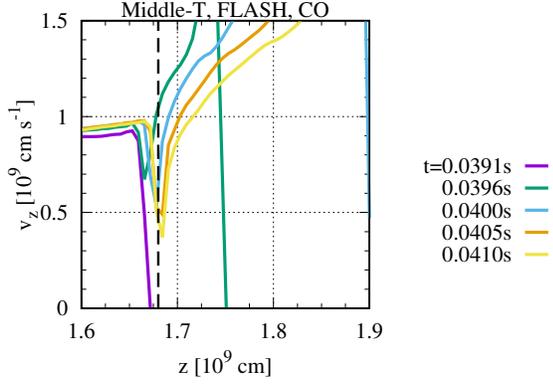}
  \caption{Enlarged figure of panel (a) of
    Figure~\ref{fig:det_extT}. \label{fig:det_extT_zoom}}
\end{figure}

Figure~\ref{fig:det_extT} shows the time evolution of $z$-direction
velocity, density, temperature, and mass fractions of nuclear elements
for the two cases after the shock wave emerges at $t \sim
0.0391$~s. For both the cases, the shock waves which emerges at $t
\sim 0.0391$~s proceed in the positive $z$-direction (see panels (a)
and (b) of Figure~\ref{fig:det_extT}). Hereafter, a shock wave
proceeding in the positive $z$-direction is referred as ``forward
shock wave''. Only for the CO case, a reverse shock wave emerges at $z
\sim 1.7 \times 10^7$~cm, which can be seen as the discontinuous
decrease of $z$-direction velocity (see Figure~\ref{fig:det_extT_zoom}
and panel (a) of Figure~\ref{fig:det_extT}). The reverse shock wave is
formed, such that materials behind the reverse shock wave experience
explosive nuclear burning (see panels (g) and (i) of
Figure~\ref{fig:det_extT}), get high temperature (see panel (e) of
Figure~\ref{fig:det_extT}), rapidly expand, and push back materials in
front of the reverse shock wave. The forward shock wave for the CO
case has higher velocity than for the ONeMg case, since the former is
energized by more active nuclear reactions than the latter.

In the following three reasons, we conclude that the reverse shock
wave accompanies a detonation wave standing at $z \sim 1.7 \times
10^7$~cm. First, upstream materials flow across $z \sim 1.7 \times
10^7$~cm at a speed of $\sim 10^9$~cm~s$^{-1}$ (see panel (a) of
Figure~\ref{fig:det_extT}), which is larger than the sound velocity of
the upstream materials ($\lesssim 5 \times 10^8$~cm~s$^{-1}$). Second,
CO elements are burned out, and Si and Fe group elements are
synthesized just behind $z \sim 1.7 \times 10^7$~cm (see panels (g)
and (i) of Figure~\ref{fig:det_extT}). Third, this fluid structure is
long-lived. It does not decay during the simulation, although we only
perform the simulation until $t \sim 0.0422$~s due to high calculation
cost of the nuclear reaction network. From $t \sim 0.0391$~s to $t
\sim 0.0422$~s, $7$~\% of materials flow across $z \sim 1.7 \times
10^7$~cm. After they experience explosive nuclear burning, they
consist of $75$~\% Fe group elements, $20$~\% Si group elements, and
$4$~\% He.

If we continue the simulation, the detonation wave will be sustained
until at least $t \sim 0.0625$~s in the following reason.  Materials
flow supersonically across $z \sim 1.7 \times 10^7$~cm from $t \sim
0.0391$~s to $t \sim 0.0625$~s in the case without considering nuclear
reactions. This should be also true in the CO case if the detonation
wave stands at $z \sim 1.7 \times 10^7$~cm during this time. This is
because upstream materials of the detonation wave should not receive
the influence of nuclear reactions due to their supersonic
motions. Then, for the CO case, materials should flow supersonically
into the detonation wave, experience explosive nuclear burning, and
sustain the detonation wave as fuels. During this time, more than
$50$~\% of materials should flow across the detonation wave. We expect
a large amount of radioactive nuclei is synthesized.

There is only the single detonation wave at $z \sim 1.7 \times
10^7$~cm; the forward shock wave at $t=0.0410$~s and $z \sim 2.3
\times 10^7$~cm does not accompany a detonation wave (see the left
panels of Figure~\ref{fig:det_extT}). We explain the reason for the
formation of only the single detonation wave at $z \sim 1.7 \times
10^7$~cm. The forward shock wave creates a hotspot at $t \sim
0.0391$~s and $t \sim 1.7 \times 10^7$~cm. The hotspot may possibly
initiates double detonation waves initially; one is the detonation
wave standing at $\sim 1.7 \times 10^7$~cm after $t \sim 0.0391$~s,
and the other may be a detonation wave which follows the forward shock
wave creating the hotspot. The former detonation wave is sustained by
high-density fuels ($\sim 3 \times 10^7$~g~cm$^{-3}$), while the
latter detonation wave may soon decay due to the supply of only
low-density fuels ($< 10^5$~g~cm$^{-3}$), as seen in panel (c) of
Figure~\ref{fig:det_extT}.

For the convergence check of a space resolution, we also perform FLASH
simulations with a space resolution of $\sim 10^6$~cm to $\sim
10^7$~cm for the middle-T region in the CO case. We find the same
detonation wave as in a space resolution of $\sim 10^5$~cm only when
the space resolution is $\lesssim 10^6$~cm. Therefore, we need
simulation with a space resolution of $\lesssim 10^6$~cm in order to
follow such a detonation wave.

For the ONeMg case, the forward shock wave raises the temperature of
materials to $> 2 \times 10^9$~K (see panel (f) of
Figure~\ref{fig:det_extT}), and triggers nuclear reactions slightly
(see panels (h) and (j) of Figure~\ref{fig:det_extT}). However, the
nuclear reactions cease soon. The forward shock wave does not excite
explosive nuclear burning triggering a reverse shock wave and a
detonation wave (see panel (b) of Figure~\ref{fig:det_extT}).

\begin{figure}[ht!]
  \plotone{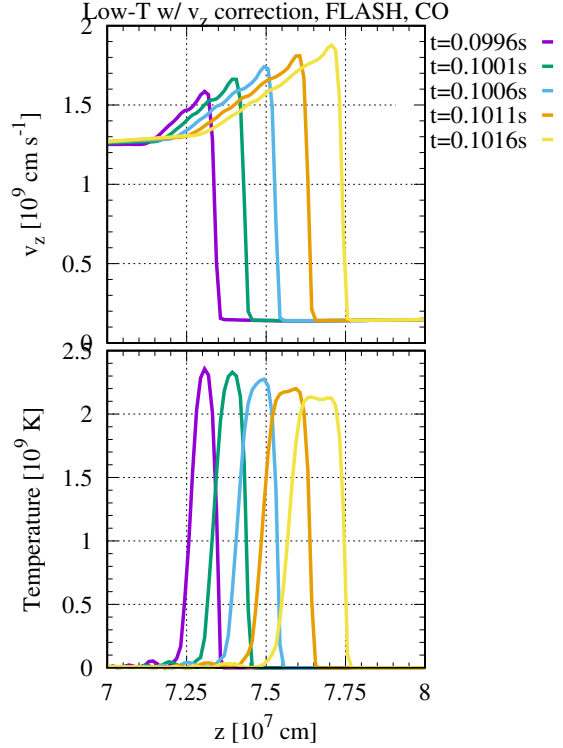}
  \caption{Time evolution of $z$-direction velocity and temperature in
    the low-T region with $v_{\textrm{z}}$ correction, when nuclear
    reactions are considered. The initial compositions are CO. The
    number of meshes is $N_{\textrm{g}}=3200$. \label{fig:det_lowT}}
\end{figure}

A forward shock wave does not always excite explosive nuclear burning
triggering a reverse shock wave and a detonation wave even when
materials consist of CO. We also perform a FLASH simulation coupled
with the nuclear reaction network for the low-T region with
$v_{\textrm{z}}$ correction in which materials consist of
CO. Figure~\ref{fig:det_lowT} shows the time evolution. At $t \sim
0.0996$~s, a forward shock wave emerges at $z \sim 7.3 \times 10^7$~cm
(see the top panel of Figure~\ref{fig:det_lowT}). The forward shock
wave raises the temperature of the materials (see the bottom panel of
Figure~\ref{fig:det_lowT}). It triggers nuclear reactions, however
does not initiate a reverse shock wave nor a detonation wave as seen
in the top panel of Figure~\ref{fig:det_lowT}. This is quite similar
to the ONeMg case in the middle-T region (see panels (b) and (f) of
Figure~\ref{fig:det_extT}).

\begin{figure}[ht!]
  \plotone{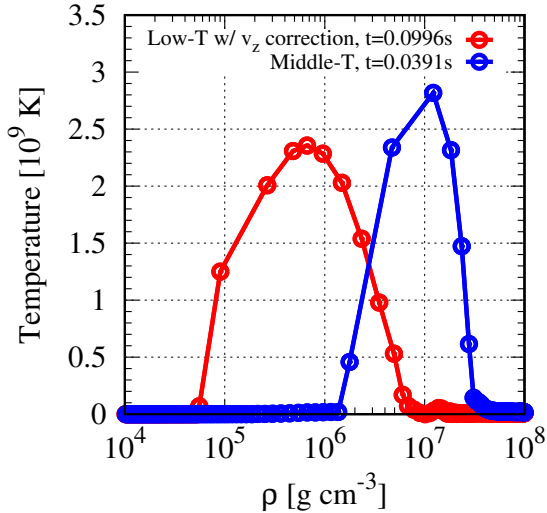}
  \caption{Density and temperature at each mesh in the low-T region
    with $v_{\textrm{z}}$ correction, and in the middle-T region, when
    shock waves emerge. \label{fig:diff_lowT-extT}}
\end{figure}

The initiation of a detonation wave strongly depends on where a
forward shock wave emerges. Figure~\ref{fig:diff_lowT-extT} shows
density and temperature in the low-T region with $v_{\textrm{z}}$
correction and in the middle-T region when forward shock waves
emerge. The forward shock wave emerges at density $\sim 10^6$ and
$\sim 10^7$~g~cm$^{-3}$ for the former and latter cases,
respectively. According to table~6 in \cite{2009ApJ...696..515S},
spontaneous initiation of a detonation wave requires a hotspot with
size of $\sim 10^7$ and $\sim 10^5$~cm for density $\sim 10^6$ and
$\sim 10^7$~g~cm$^{-3}$, respectively. Since a mesh size is $\sim
10^5$~cm, the hotspot is too small to generate a detonation wave for
the former case, and is sufficiently large for the latter case. Our
results about the initiation of the detonation wave are consistent
with the criteria in \cite{2009ApJ...696..515S}.

As seen in table~8, 9, 10, and 11 of \cite{2009ApJ...696..515S}, the
hotspot size to generate a detonation wave becomes rapidly larger as
carbon fraction becomes smaller. Even when the carbon fraction
decreases from $50$~\% to $30$~\%, a hotspot size required to generate
a detonation wave becomes larger by a factor of $\sim 10$. Therefore,
the reason why no detonation wave occurs for the ONeMg case in the
middle-T region (see the right panels of Figure~\ref{fig:det_extT}) is
that a hotspot is too small to generate a detonation wave in ONeMg
compositions.

From the above, we find two implications. First, the resolution of 3D
SPH simulations is too low to resolve a forward shock wave even if $N
> 10^7$. Second, the forward shock wave does not always initiate a
detonation wave even if materials consist of CO.

Finally, we discuss the reliability of the initiation of a detonation
wave for the CO case in the middle-T region. In this case, the forward
shock wave occurs, and create the hotspot generating the detonation
wave where the density gradient is steep. As seen in
Figure~\ref{fig:diff_lowT-extT}, the hotspot has density of
$10^7$~g~cm$^{-3}$, however a mesh near the hotspot has density of
$\sim 10^6$~g~cm$^{-3}$. If the forward shock wave emerges at a bit
outer region, it can not trigger a detonation wave, since the forward
shock wave propagates outward (see panel (a) of
Figure~\ref{fig:det_extT}). Our 1D initial conditions lack accuracy to
precisely determine where a forward shock wave occurs. Therefore, the
initiation of a detonation wave may be unreliable. In order to follow
the initiation of a detonation wave accurately, we should perform 3D
simulations with a space resolution of $\lesssim 10^6$~cm. Note that
the detonation wave in the middle-T region occurs only when a space
resolution is $\lesssim 10^6$~cm.

\section{Discussion}
\label{sec:discussion}

In this section, we discuss why materials are falsely heated in 3D SPH
simulation with low resolution, using model CO1 which is similar to
model of Rosswog's run~8.

Model CO1 provides a solid basis for the comparison, since this
parameter set is the same as that of Rosswog's run~8, and
\cite{2016ApJ...819....3M} have investigated the observational
features of this model. In our model, the WDs yield $0.13M_\odot$ and
$0.15M_\odot$ of Fe group elements for $N=0.39$M and $N=0.79$M,
respectively. On the other hand, Rosswog's run~8 has $0.18M_\odot$ for
$N=0.5$M. Our results are consistent with those of Rosswog's run~8
from the view point of the nucleosynthesis.

\begin{figure*}[ht!]
  \plotone{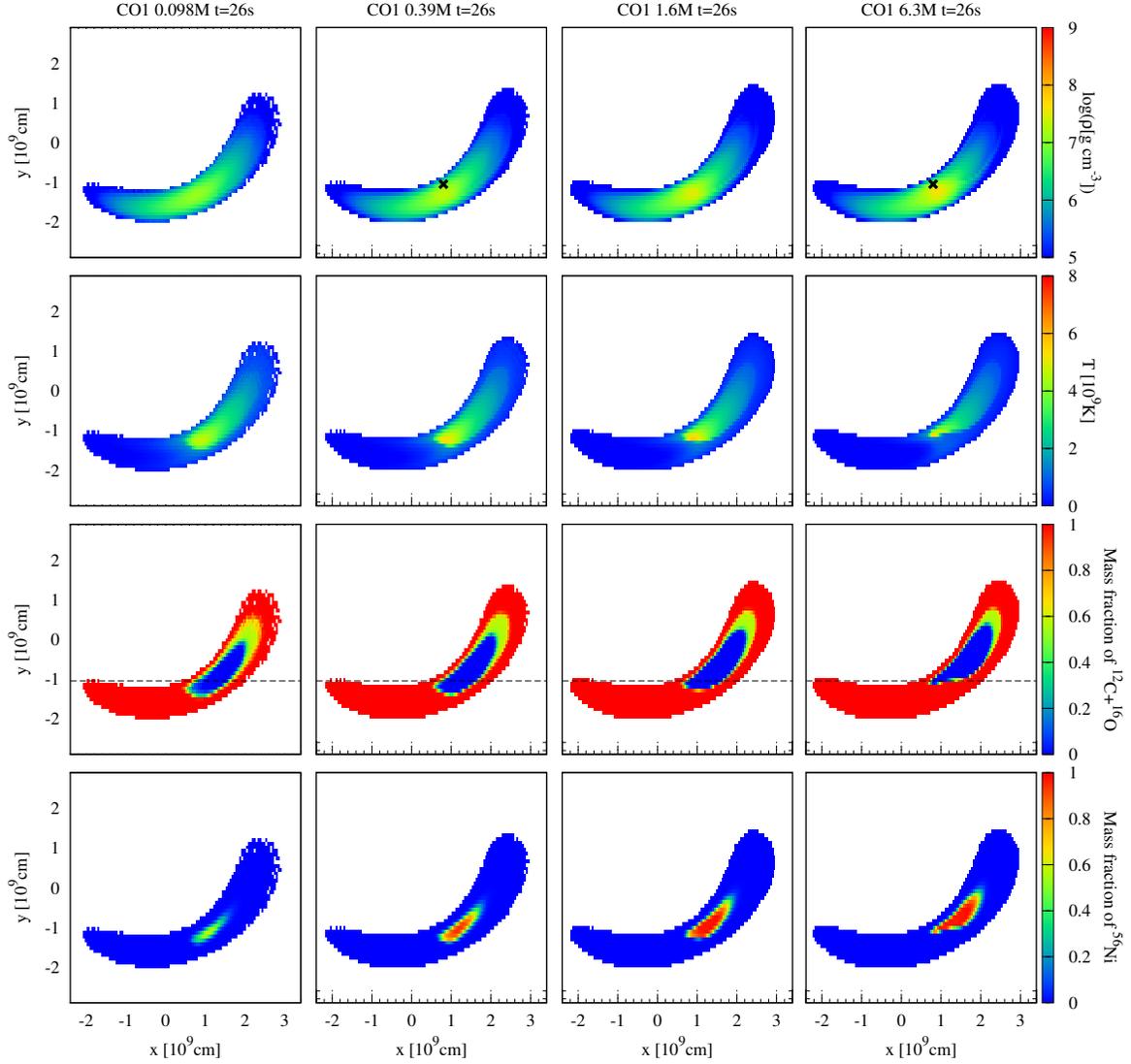}
  \caption{Density, temperature, and mass fractions of original
    elements and $^{56}$Ni on the orbital plane in model CO1 at
    $t=26$~s, when nuclear reactions are going on. From left to right,
    $N$ is $0.098$M, $0.39$M, $1.6$M, and $6.3$M. The dashed lines
    indicate cross sections drawn in Figure~\ref{fig:xzplane_co1}. The
    crosses denote particles investigated in detail in
    Figures~\ref{fig:dens-temp_co1}, \ref{fig:x-z_co1}, and
    \ref{fig:x-vz_co1}. \label{fig:xyplane_co1}}
\end{figure*}

Figure~\ref{fig:xyplane_co1} shows thermodynamical quantities of model
CO1 in the same way as in Figure~\ref{fig:xyplane_onemg}. As seen in
the bottom panels, the amount of $^{56}$Ni increases from $N=0.098$M
to $N=0.39$M, and decreases from $N=1.6$M to $N=6.3$M, which is
consistent with the amount of $^{56}$Ni drawn in
Figure~\ref{fig:n-element}. The increase from $N=0.098$M to $N=0.39$M
comes from the increase of the density of particles with increasing
$N$ (see the top panels). Note that $^{56}$Ni is more synthesized
(while artificially) at higher density if materials experience
explosive nuclear burning. The low density in $N=0.098$M results in
the lower temperature and larger unburned mass in $N=0.098$M than
those in $N=0.39$M. Since the density is low, the nuclear reactions
are not active so much.

The dependence of the density on $N$ can be explained as follows. As
$N$ becomes smaller, a particle on the orbital plane has a larger
kernel-support radius. SPH simulation calculates the density of the
particle, considering more distant particles from the orbital
plane. Particles are distributed more sparsely with increasing
distance from the orbital plane. Therefore, the density of the
particle becomes smaller with a larger kernel-support radius (smaller
$N$).

\begin{figure*}[ht!]
  \plotone{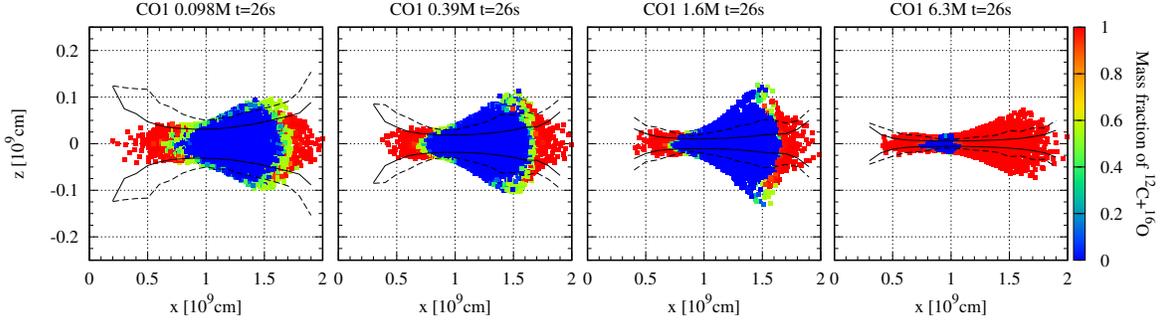}
  \caption{Mass fractions of original elements on a plane
    perpendicular to the orbital plane in model CO1 at $t=26$~s. The
    plane is indicated by the dashed lines in
    Figure~\ref{fig:xyplane_co1}. The width of two black solid
    (dashed) curves in the $z$-direction is equal to twice the minimum
    (maximum) value of the kernel-support radii among the particles at
    the $x$-coordinate. From left to right, $N$ is $0.098$M, $0.39$M,
    $1.6$M, and $6.3$M. \label{fig:xzplane_co1}}
\end{figure*}

The decrease of $^{56}$Ni from $N=1.6$M to $N=6.3$M is also due to the
same reason as the decrease in model
ONeMg. Figure~\ref{fig:xzplane_co1} shows the structure of model CO1
on a plane perpendicular to the orbital plane. In $N=6.3$M, the
kernel-support radii of the particles are smaller than the scale
height, except around $x=10^9$~cm where explosive nuclear burning
occurs. In $N=0.098$M, $0.39$M, and $1.6$M, the kernel-support radii
seem smaller than the scale height, even where explosive nuclear
burning occurs. However, the kernel-support radius is comparable to
the scale height at the time earlier than $t=26$~s.

In the following, we clarify the reason why the nuclear reactions
occur if the kernel-support radius is comparable to the scale height
in the $z$-direction. For this purpose, we follow the time evolution
of particles located at the crosses in Figure~\ref{fig:xyplane_co1}. In
order to focus on whether the nuclear reactions start or not, we adopt
model CO1 w/o nuc.

\begin{figure}[ht!]
  \plotone{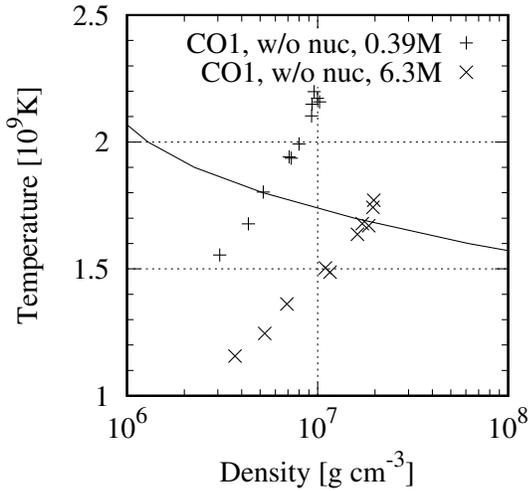}
  \caption{Density and temperature of particles indicated in
    Figure~\ref{fig:xyplane_co1}. These particles are in the
    $z$-direction column centered on $(x,y)=(0.8 \times
    10^9\mbox{cm},-1.05 \times 10^9\mbox{cm})$. The width of the
    column is $10^7$~cm for $N=0.39$M, and $2\times 10^6$~cm for
    $N=6.3$M. The black curve shows the threshold for explosive
    nuclear burning. On the curve, the $^{12}$C + $^{12}$C reaction
    timescale is equal to the local dynamical timescale. Both the
    timescale are defined in the same way as
    \cite{2015ApJ...807...40T}. If a particle is on the upper side
    from the curve, the particle experiences the explosive nuclear
    burning. \label{fig:dens-temp_co1}}
\end{figure}

Figure~\ref{fig:dens-temp_co1} shows the density and temperature of
these particles in $N=0.39$M and $6.3$M. Since the $^{12}$C + $^{12}$C
timescale is shorter than the local dynamical timescale in a part of
particles in both $N$, they would experience the explosive nuclear
burning if the nuclear reaction network was turned on.

\begin{figure}[ht!]
  \plotone{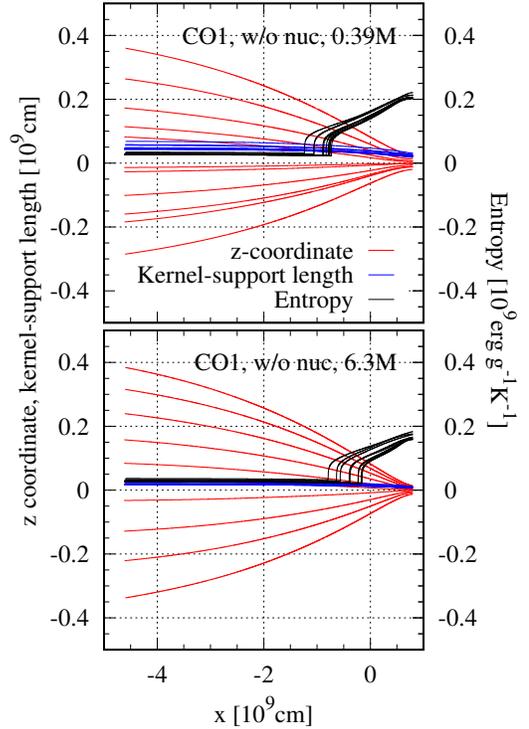}
  \caption{Trajectories, kernel-support radii, and entropy of the same
    particles as those in Figure~\ref{fig:dens-temp_co1}. These
    particles move from left to right with time. \label{fig:x-z_co1}}
\end{figure}

We can see the trajectories of these particles in
Figure~\ref{fig:x-z_co1}. The trajectories are independent of $N$ from
$x=-4.5 \times 10^9$~cm to $x=0$~cm. The particles are in the range of
$z=\pm 0.4 \times 10^9$~cm at $x=-4.5 \times 10^9$~cm, and in the
range of $z=\pm 0.1 \times 10^9$~cm at $x=0$~cm. At $x=0.8 \times
10^9$~cm, the particles in $N=6.3$M are distributed slightly more
densely than those in $N=0.39$M. Since these particles approach the
orbital plane, the scale height of the structure shrinks in the
$z$-direction. On the other hand, the kernel-support radii of these
particles keep nearly constant; the density of the structure does not
grow. This is because the structure is compressed in the
$z$-direction, but is extended in the direction of the orbital plane.

The entropy of these particles grows from some point. Although their
entropy increases, no shock wave is found out. If there is a shock
wave, the trajectories of these particles should be changed
discontinuously. Moreover, their entropy in $N=0.39$M grows at smaller
$x$ (i.e. earlier) than those in $N=6.3$M. These behaviours of their
entropy are clearly not converged against $N$.

\begin{figure}[ht!]
  \plotone{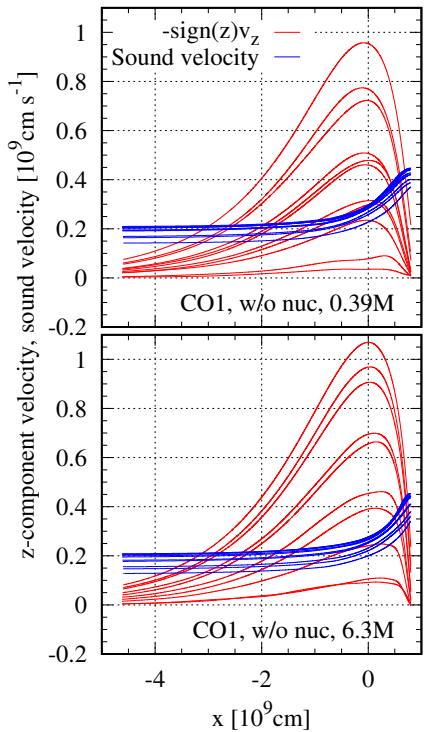}
  \caption{Velocities in the $z$-direction, and sound velocities of
    the same particles as those in Figure~\ref{fig:dens-temp_co1}. The
    velocities in the $z$-direction are defined as velocities whose
    signs are positive (negative) if the particles approach (recede
    from) the orbital plane. These particles move from left to right
    with time. \label{fig:x-vz_co1}}
\end{figure}

We investigate why the entropy increases despite the absence of a
shock wave. Figure~\ref{fig:x-vz_co1} shows the velocities in the
$z$-direction and sound velocities of these particles. Particles
distant from the orbital plane approach the orbital plane
supersonically from $x=-3\times 10^9$~cm to $x=0.8 \times 10^9$~cm,
while particles near the orbital plane have small velocities in the
$z$-direction. Therefore, the relative velocities between the
outermost and innermost particles from the orbital plane are
supersonic. Since all these particles have the kernel-support radii
comparable to the scale height, the innermost particles interact
supersonically with the outermost particles. Then, the particles
obtain entropy.

The reason why particles in smaller $N$ obtain entropy earlier is as
follows. The kernel-support radii of particles are larger in smaller
$N$. Therefore, in smaller $N$, the innermost particles start
interacting with the outermost particles when the scale height is
still larger.

As $N$ becomes larger, the kernel-support radii become smaller. Each
of the innermost and outermost particles interacts only with their
adjacent particles. Their relative velocities are
subsonic. Consequently, these particles do not gain entropy in large
$N$.

In summary, particles on the orbital plane are heated and gain
entropy, interacting with the outermost particles from the orbital
plane due to their kernel-support radii comparable to the scale
height. Such interactions should not happen in reality, and the
heating is spurious. The spurious heating triggers explosive nuclear
burning falsely. Eventually, our 3D SPH simulations fail to follow the
nucleosynthesis in WD~TDEs, even if our simulations have
unprecedentedly high resolution. This should also be the case for
previous simulations with lower resolution.

\section{Summary}
\label{sec:summary}

We perform 3D SPH simulations with $N>10^7$ to follow the TDEs of
He~WDs, CO~WDs, and ONeMg~WDs by IMBHs. We observe that the explosive
nuclear burning occurs in all the WDs. However, the final compositions
are strongly dependent on $N$. We find that the amount of unburned
materials increases with increasing $N$. The nucleosynthesis of these
WDs is not converged against $N$, although our simulations contain
unprecedentedly large $N$, up to a few $10^7$ particles.

The reason why the explosive nuclear burning occurs in small-$N$
simulations is as follows. In such small-$N$ simulations, the scale
height of a WD in the $z$-direction becomes comparable to the
kernel-support radii of particles at the pericenter passage around an
IMBH. On the other hand, a particle at the surface of the WD
approaches the orbital plane supersonically. Then, the particle
interacts with other particles on the orbital plane. Their
interactions generate spurious heating, and trigger explosive nuclear
burning falsely. If $N$ is sufficiently large, these particles do not
interact with each other, since the kernel-support radius is
sufficiently small.

By means of mesh-based simulations, \cite{2012ApJ...749..117H} have
suggested that a WD~TDE can experience explosive nuclear
burning. However, we conjecture that their mesh-based simulations have
not either resolved the scale height of the WD in the $z$-direction at
the pericenter passage. They have investigated TDEs with $0.6M_\odot$
CO~WDs which orbit around a $10^3M_\odot$ IMBH with $\beta=6$. This
model is similar to our model CO. In our model, the scale height in
the $z$-direction is about $2 \times 10^7$~cm at the pericenter
passage when $\beta = 5$. If the scale height is almost equal to those
in \cite{2012ApJ...749..117H}, their simulations can not resolve the
scale height, since their finest mesh size is about $10^7$~cm.

We find no shock wave generation in 3D SPH simulations with $N >
10^7$. So, we perform 1D SPH and FLASH simulations with higher
resolution than the 3D SPH simulations in order to examine whether the
absence of shock waves results from the low resolution of the 3D SPH
simulations. Eventually, we find the following two insights. First, a
shock wave emerges in both 1D SPH and FLASH simulations. In other
words, the resolution of the current 3D SPH simulations is not enough
to resolve the shock wave, even if $N > 10^7$. Second, the shock wave
can trigger a detonation wave, only when it makes a hotspot large
enough to generate a detonation wave. For the case of materials
consisting of ONeMg, no shock wave triggers a detonation. For the case
of materials consisting of CO, if the shock wave occurs in a
low-density region ($\lesssim 10^6$~g~cm$^{-3}$), a hotspot formed by
the shock wave is too small to generate a detonation wave.

Although a detonation wave occurs in a part of our 1D simulations, the
initiation of the detonation wave may be unreliable for the following
reason. The shock wave triggering the detonation wave occurs where the
density gradient is steep. If the shock wave occurs at a bit outer
region, it can not trigger the detonation wave, since it occurs in
much lower density region by a factor of $10$. Our 1D modeling lacks
accuracy to precisely determine where the shock wave occurs.

In order to conclude that WD~TDEs become optical transients powered by
radioactive nuclei, we need to perform 3D simulations with a space
resolution of $\lesssim 10^6$~cm. SPH simulation with $N \sim 10^9$
seems to satisfy this requirement. However, this may not be true in
the following reason. A shock wave emerges at the surface of a
WD. Generally, SPH simulation has lower space resolution at the
surface of an object than inside. Therefore, SPH simulation may need
$N \gg 10^9$ particles.

\acknowledgments

A. Tanikawa thanks K. Kawana, N. Yoshida, Y. Guo, and K. Sugiura for
fruitful discussions. Numerical computations were carried out on Cray
XC30 at Center for Computational Astrophysics, National Astronomical
Observatory of Japan, and on Cray XC40 at Yukawa Institute for
Theoretical Physics, Kyoto University. The software used in this work
was in part developed by the DOE NNSA-ASC OASCR Flash Center at the
University of Chicago. This research has been supported in part by
World Premier International Research Center Initiative (WPI
Initiative), MEXT, Japan, by MEXT program for the Development and
Improvement for the Next Generation Ultra High-Speed Computer System
under its Subsidies for Operating the Specific Advanced Large Research
Facilities, and by Grants-in-Aid for Scientific Research (24540227,
26400222, 26800100, 16H02168, 16K17656) from the Japan Society for the
Promotion of Science. Finally, we thank the anonymous referee for
helpful advice.


\begin{thebibliography}{}
\expandafter\ifx\csname natexlab\endcsname\relax\def\natexlab#1{#1}\fi

\bibitem[{{Balsara}(1995)}]{1995JCoPh.121..357B}
{Balsara}, D.~S. 1995, Journal of Computational Physics, 121, 357

\bibitem[{{Brassart} \& {Luminet}(2008)}]{2008A&A...481..259B}
{Brassart}, M., \& {Luminet}, J.-P. 2008, \aap, 481, 259

\bibitem[{{Cheng} \& {Bogdanovi{\'c}}(2014)}]{2014PhRvD..90f4020C}
{Cheng}, R.~M., \& {Bogdanovi{\'c}}, T. 2014, \prd, 90, 064020

\bibitem[{{Clausen} \& {Eracleous}(2011)}]{2011ApJ...726...34C}
{Clausen}, D., \& {Eracleous}, M. 2011, \apj, 726, 34

\bibitem[{{Colella} \& {Woodward}(1984)}]{1984JCoPh..54..174C}
{Colella}, P., \& {Woodward}, P.~R. 1984, Journal of Computational Physics, 54,
  174

\bibitem[{{Dehnen} \& {Aly}(2012)}]{2012MNRAS.425.1068D}
{Dehnen}, W., \& {Aly}, H. 2012, \mnras, 425, 1068

\bibitem[{{East}(2014)}]{2014ApJ...795..135E}
{East}, W.~E. 2014, \apj, 795, 135

\bibitem[{{Farrell} {et~al.}(2009){Farrell}, {Webb}, {Barret}, {Godet}, \&
  {Rodrigues}}]{2009Natur.460...73F}
{Farrell}, S.~A., {Webb}, N.~A., {Barret}, D., {Godet}, O., \& {Rodrigues},
  J.~M. 2009, \nat, 460, 73

\bibitem[{{Fryxell} {et~al.}(2000){Fryxell}, {Olson}, {Ricker}, {Timmes},
  {Zingale}, {Lamb}, {MacNeice}, {Rosner}, {Truran}, \&
  {Tufo}}]{2000ApJS..131..273F}
{Fryxell}, B., {Olson}, K., {Ricker}, P., {et~al.} 2000, \apjs, 131, 273

\bibitem[{{Haas} {et~al.}(2012){Haas}, {Shcherbakov}, {Bode}, \&
  {Laguna}}]{2012ApJ...749..117H}
{Haas}, R., {Shcherbakov}, R.~V., {Bode}, T., \& {Laguna}, P. 2012, \apj, 749,
  117

\bibitem[{{Ioka} {et~al.}(2016){Ioka}, {Hotokezaka}, \&
  {Piran}}]{2016ApJ...833..110I}
{Ioka}, K., {Hotokezaka}, K., \& {Piran}, T. 2016, \apj, 833, 110

\bibitem[{Iwasawa {et~al.}(2015)Iwasawa, Tanikawa, Hosono, Nitadori, Muranushi,
  \& Makino}]{Iwasawa:2015:FNF:2830018.2830019}
Iwasawa, M., Tanikawa, A., Hosono, N., {et~al.} 2015, in Proceedings of the 5th
  International Workshop on Domain-Specific Languages and High-Level Frameworks
  for High Performance Computing, WOLFHPC '15 (New York, NY, USA: ACM),
  1:1--1:10

\bibitem[{{Iwasawa} {et~al.}(2016){Iwasawa}, {Tanikawa}, {Hosono}, {Nitadori},
  {Muranushi}, \& {Makino}}]{2016PASJ...68...54I}
{Iwasawa}, M., {Tanikawa}, A., {Hosono}, N., {et~al.} 2016, \pasj, 68, 54

\bibitem[{{Jonker} {et~al.}(2013){Jonker}, {Glennie}, {Heida}, {Maccarone},
  {Hodgkin}, {Nelemans}, {Miller-Jones}, {Torres}, \&
  {Fender}}]{2013ApJ...779...14J}
{Jonker}, P.~G., {Glennie}, A., {Heida}, M., {et~al.} 2013, \apj, 779, 14

\bibitem[{{Kobayashi} {et~al.}(2004){Kobayashi}, {Laguna}, {Phinney}, \&
  {M{\'e}sz{\'a}ros}}]{2004ApJ...615..855K}
{Kobayashi}, S., {Laguna}, P., {Phinney}, E.~S., \& {M{\'e}sz{\'a}ros}, P.
  2004, \apj, 615, 855

\bibitem[{{Komossa}(2015)}]{2015JHEAp...7..148K}
{Komossa}, S. 2015, Journal of High Energy Astrophysics, 7, 148

\bibitem[{{Krolik} \& {Piran}(2011)}]{2011ApJ...743..134K}
{Krolik}, J.~H., \& {Piran}, T. 2011, \apj, 743, 134

\bibitem[{{Luminet} \& {Carter}(1986)}]{1986ApJS...61..219L}
{Luminet}, J.-P., \& {Carter}, B. 1986, \apjs, 61, 219

\bibitem[{{Luminet} \& {Pichon}(1989)}]{1989A&A...209..103L}
{Luminet}, J.-P., \& {Pichon}, B. 1989, \aap, 209, 103

\bibitem[{{MacLeod} {et~al.}(2014){MacLeod}, {Goldstein}, {Ramirez-Ruiz},
  {Guillochon}, \& {Samsing}}]{2014ApJ...794....9M}
{MacLeod}, M., {Goldstein}, J., {Ramirez-Ruiz}, E., {Guillochon}, J., \&
  {Samsing}, J. 2014, \apj, 794, 9

\bibitem[{{MacLeod} {et~al.}(2016){MacLeod}, {Guillochon}, {Ramirez-Ruiz},
  {Kasen}, \& {Rosswog}}]{2016ApJ...819....3M}
{MacLeod}, M., {Guillochon}, J., {Ramirez-Ruiz}, E., {Kasen}, D., \& {Rosswog},
  S. 2016, \apj, 819, 3

\bibitem[{{Matsumoto} {et~al.}(2001){Matsumoto}, {Tsuru}, {Koyama}, {Awaki},
  {Canizares}, {Kawai}, {Matsushita}, \& {Kawabe}}]{2001ApJ...547L..25M}
{Matsumoto}, H., {Tsuru}, T.~G., {Koyama}, K., {et~al.} 2001, \apj, 547, L25

\bibitem[{{Monaghan}(1997)}]{1997JCoPh.136..298M}
{Monaghan}, J.~J. 1997, Journal of Computational Physics, 136, 298

\bibitem[{{Morris} \& {Monaghan}(1997)}]{1997JCoPh.136...41M}
{Morris}, J.~P., \& {Monaghan}, J.~J. 1997, Journal of Computational Physics,
  136, 41

\bibitem[{{Paczy{\'n}sky} \& {Wiita}(1980)}]{1980A&A....88...23P}
{Paczy{\'n}sky}, B., \& {Wiita}, P.~J. 1980, \aap, 88, 23

\bibitem[{{Price} \& {Monaghan}(2007)}]{2007MNRAS.374.1347P}
{Price}, D.~J., \& {Monaghan}, J.~J. 2007, \mnras, 374, 1347

\bibitem[{{Raskin} {et~al.}(2010){Raskin}, {Scannapieco}, {Rockefeller},
  {Fryer}, {Diehl}, \& {Timmes}}]{2010ApJ...724..111R}
{Raskin}, C., {Scannapieco}, E., {Rockefeller}, G., {et~al.} 2010, \apj, 724,
  111

\bibitem[{{Rees}(1984)}]{1984ARA&A..22..471R}
{Rees}, M.~J. 1984, \araa, 22, 471

\bibitem[{{Rosswog}(2005)}]{2005ApJ...634.1202R}
{Rosswog}, S. 2005, \apj, 634, 1202

\bibitem[{{Rosswog} {et~al.}(2009){Rosswog}, {Ramirez-Ruiz}, \&
  {Hix}}]{2009ApJ...695..404R}
{Rosswog}, S., {Ramirez-Ruiz}, E., \& {Hix}, W.~R. 2009, \apj, 695, 404

\bibitem[{{Rosswog} {et~al.}(2008){Rosswog}, {Ramirez-Ruiz}, {Hix}, \&
  {Dan}}]{2008CoPhC.179..184R}
{Rosswog}, S., {Ramirez-Ruiz}, E., {Hix}, W.~R., \& {Dan}, M. 2008, Computer
  Physics Communications, 179, 184

\bibitem[{{Sato} {et~al.}(2015){Sato}, {Nakasato}, {Tanikawa}, {Nomoto},
  {Maeda}, \& {Hachisu}}]{2015ApJ...807..105S}
{Sato}, Y., {Nakasato}, N., {Tanikawa}, A., {et~al.} 2015, \apj, 807, 105

\bibitem[{{Sato} {et~al.}(2016){Sato}, {Nakasato}, {Tanikawa}, {Nomoto},
  {Maeda}, \& {Hachisu}}]{2016ApJ...821...67S}
---. 2016, \apj, 821, 67

\bibitem[{{Seitenzahl} {et~al.}(2009){Seitenzahl}, {Meakin}, {Townsley},
  {Lamb}, \& {Truran}}]{2009ApJ...696..515S}
{Seitenzahl}, I.~R., {Meakin}, C.~A., {Townsley}, D.~M., {Lamb}, D.~Q., \&
  {Truran}, J.~W. 2009, \apj, 696, 515

\bibitem[{{Sell} {et~al.}(2015){Sell}, {Maccarone}, {Kotak}, {Knigge}, \&
  {Sand}}]{2015MNRAS.450.4198S}
{Sell}, P.~H., {Maccarone}, T.~J., {Kotak}, R., {Knigge}, C., \& {Sand}, D.~J.
  2015, \mnras, 450, 4198

\bibitem[{{Shcherbakov} {et~al.}(2013){Shcherbakov}, {Pe'er}, {Reynolds},
  {Haas}, {Bode}, \& {Laguna}}]{2013ApJ...769...85S}
{Shcherbakov}, R.~V., {Pe'er}, A., {Reynolds}, C.~S., {et~al.} 2013, \apj, 769,
  85

\bibitem[{{Shiokawa} {et~al.}(2015){Shiokawa}, {Krolik}, {Cheng}, {Piran}, \&
  {Noble}}]{2015ApJ...804...85S}
{Shiokawa}, H., {Krolik}, J.~H., {Cheng}, R.~M., {Piran}, T., \& {Noble}, S.~C.
  2015, \apj, 804, 85

\bibitem[{{Stone} {et~al.}(2013){Stone}, {Sari}, \&
  {Loeb}}]{2013MNRAS.435.1809S}
{Stone}, N., {Sari}, R., \& {Loeb}, A. 2013, \mnras, 435, 1809

\bibitem[{{Tanikawa} {et~al.}(2015){Tanikawa}, {Nakasato}, {Sato}, {Nomoto},
  {Maeda}, \& {Hachisu}}]{2015ApJ...807...40T}
{Tanikawa}, A., {Nakasato}, N., {Sato}, Y., {et~al.} 2015, \apj, 807, 40

\bibitem[{{Tanikawa} {et~al.}(2013){Tanikawa}, {Yoshikawa}, {Nitadori}, \&
  {Okamoto}}]{2013NewA...19...74T}
{Tanikawa}, A., {Yoshikawa}, K., {Nitadori}, K., \& {Okamoto}, T. 2013, \na,
  19, 74

\bibitem[{{Tanikawa} {et~al.}(2012){Tanikawa}, {Yoshikawa}, {Okamoto}, \&
  {Nitadori}}]{2012NewA...17...82T}
{Tanikawa}, A., {Yoshikawa}, K., {Okamoto}, T., \& {Nitadori}, K. 2012, \na,
  17, 82

\bibitem[{{Tejeda} {et~al.}(2017){Tejeda}, {Gafton}, \&
  {Rosswog}}]{2017arXiv170100303T}
{Tejeda}, E., {Gafton}, E., \& {Rosswog}, S. 2017, ArXiv e-prints,
  arXiv:1701.00303

\bibitem[{{Tejeda} \& {Rosswog}(2013)}]{2013MNRAS.433.1930T}
{Tejeda}, E., \& {Rosswog}, S. 2013, \mnras, 433, 1930

\bibitem[{{Timmes} {et~al.}(2000){Timmes}, {Hoffman}, \&
  {Woosley}}]{2000ApJS..129..377T}
{Timmes}, F.~X., {Hoffman}, R.~D., \& {Woosley}, S.~E. 2000, \apjs, 129, 377

\bibitem[{{Timmes} \& {Swesty}(2000)}]{2000ApJS..126..501T}
{Timmes}, F.~X., \& {Swesty}, F.~D. 2000, \apjs, 126, 501

\bibitem[{Wendland(1995)}]{wendland1995piecewise}
Wendland, H. 1995, Advances in Computational Mathematics, 4, 389

\bibitem[{{Zalamea} {et~al.}(2010){Zalamea}, {Menou}, \&
  {Beloborodov}}]{2010MNRAS.409L..25Z}
{Zalamea}, I., {Menou}, K., \& {Beloborodov}, A.~M. 2010, \mnras, 409, L25

\end{thebibliography}

\end{document}